\documentclass[runningheads]{llncs}
\usepackage[T1]{fontenc}
\usepackage{graphicx}
\usepackage{hyperref}
\usepackage{url} 
\usepackage{hyperref}
\usepackage{cite}
\usepackage{amsmath}
\usepackage{booktabs} 
\usepackage{graphicx}
\graphicspath{ {figures/} }
\usepackage{graphicx}
\usepackage{cite} 
\usepackage{listings}
\begin{document}
\title{RAMBO: Leaking Secrets from Air-Gap Computers by Spelling Covert Radio Signals from Computer RAM}
%
%
\author{Mordechai Guri\orcidID{0000-0003-1806-8858}}
\authorrunning{M. Guri}
\institute{Ben-Gurion University of the Negev, Israel\\ Department of Software and Information Systems Engineering 
	\email{gurim@post.bgu.ac.il}\\ Air-gap research page:
	\texttt{http://www.covertchannels.com}}

%

\maketitle              
\begin{abstract}
Air-gapped systems are physically separated from external networks, including the Internet. This isolation is achieved by keeping the air-gap computers disconnected from wired or wireless networks, preventing direct or remote communication with other devices or networks. Air-gap measures may be used in sensitive environments where security and isolation are critical to prevent private and confidential information leakage. 

In this paper, we present an attack allowing adversaries to leak information from air-gapped computers. We show that malware on a compromised computer can generate radio signals from memory buses (RAM). Using software-generated radio signals, malware can encode sensitive information such as files, images, keylogging, biometric information, and encryption keys. With software-defined radio (SDR) hardware, and a simple off-the-shelf antenna, an attacker can intercept transmitted raw radio signals from a distance. The signals can then be decoded and translated back into binary information. We discuss the design and implementation and present related work and evaluation results. This paper presents fast modification methods to leak data from air-gapped computers at 1000 bits per second. Finally, we propose countermeasures to mitigate this out-of-band air-gap threat.

\keywords{Air-gap \and Radio \and Electromagnetic \and Covert Channels \and Exfiltration \and RAM \and Memory}
\end{abstract}
\section{Introduction}
\label{sec:intro}
Today’s regulations, such as GDPR (General Data Protection Regulation), outline strict rules and principles for how organizations should collect, store, and share personal data. It grants individuals certain rights, such as the right to access their data, the right to be forgotten (i.e., to have their data erased), the right to data portability, and more. Organizations that handle personal data must follow certain practices to ensure privacy and security. They need explicit consent from individuals before processing their data. They need to implement strong data protection measures and report data breaches within a specific timeframe \cite{albrecht2016gdpr}.

When sensitive data such as personal or confidential information is involved, the collection, processing, and storage of the information may be done in networks disconnected from the Internet. This security measure is known as an  'air gap.' Air-gap isolation protects information from cyberattacks, and online risks, including phishing emails, social engineering, and compromised websites \cite{guri2018bridgeware}.
\subsection{Air-gap Isolation}
Enforcing an air gap in a computing or networking environment involves physically and logically isolating a system, network, or device from external networks or communication channels. This can be done by disconnecting network cables, disabling wireless interfaces, and disallowing USB connections. In addition, it must be ensured that the isolated system has no direct link to any external communication infrastructure \cite{guri2021usbculprit}.
\subsection{Air-gap Attacks}
Despite air-gapped networks being considered highly secure, there have been incidents demonstrating that air-gapped networks are not immune to breaches. Stuxnet is one of the most famous air-gap malware \cite{chen2011lessons}. Discovered in 2010, Stuxnet was a highly sophisticated worm that targeted industrial control systems (ICS), particularly those used in nuclear facilities. It exploited zero-day vulnerabilities and used several methods, including infected USB drives, to jump the air gap and spread it across isolated networks. The Agent.BTZ worm \cite{gostev2014agent} was another type of air gap computer worm with advanced capabilities and a targeted type. It was specifically designed to spread through removable media, such as USB flash drives, and infiltrate computer networks, including highly secure or air-gapped. According to reports, the worm affected the U.S. Department of Defense classified networks. Notably more than twenty-five reported malware in the past targeted highly secured and air-gapped networks \cite{dorais2021jumping}, including USBStealer, Agent.BTZ \cite{gostev2014agent}, Stuxnet \cite{chen2011lessons}, Fanny, MiniFlame, Flame, Gauss, ProjectSauron, EZCheese, Emotional Simian, USB Thief, USBFerry, Retro, and Ramsay.
\subsection{The RAMBO Attack}
In order to exfiltrate information from an infected air-gapped computer, attackers use special communication channels known as air-gap covert channels. There are several types of covert channels studied in the past twenty years \cite{cabaj2018new}\cite{caviglione2021trends}. These attacks leak data through electromagnetic emission \cite{Guri2015}\cite{Guri2014}\cite{shen2021lora}, optical signals \cite{guri2019ctrl}, acoustic noise \cite{guri2020fansmitter}\cite{guri2022gpu}, thermal changes \cite{guri2015bitwhisper}, and even physical vibrations \cite{guri2021exfiltrating}.
In this paper, we show how malware can manipulate RAM to generate radio signals at clock frequencies. These signals are modified and encoded in a particular encoding allowing them to be received from a distance away. The attacker can encode sensitive information (keylogging, documents, images, biometric information, etc.) and exfiltrate it via these radio signals. An attacker with appropriate hardware can receive the electromagnetic signals, demodulate and decode the data, and retrieve the exfiltrated information.

This paper is organized as follows. The attack model is first described in Section \ref{sec:attack}. Section \ref{sec:related} provides a review of related work. Section \ref{sec:trans} describes the design and implementation of a transmitter and receiver, including modulation and encoding. The analysis and evaluation results are presented in section \ref{sec:eval}. Section \ref{sec:counter} provides a list of countermeasures, and we conclude in Section \ref{sec:conclusion}.

\section{Attack Model}
\label{sec:attack}
Attacks on air-gapped networks involve multi-phase strategies to breach isolated systems by delivering specialized malware through physical media or insider agents, initiating malware execution, propagating within the network, exfiltrating data using covert channels or compromised removable media, establishing remote command and control, evading detection, and covering tracks. In the context of the RAMBO attack, the adversary must infect the air-gap network in the initial phase. This can be done via a variety of attack vectors \cite{dorais2021jumping}\cite{AirGappe49:online}\cite{Beatingt3:online}\cite{Guri2018b}.

An attacker could plant malware on a USB drive and physically introduce it into an air-gapped network. An unsuspecting insider or employee might connect the USB drive to a computer within the isolated network, unknowingly activating the malware and allowing it to propagate and exfiltrate data through the same USB drive or via covert channels. An insider with access to the air-gapped network might intentionally introduce malware or provide unauthorized access to external parties. This could involve transferring sensitive data to personal devices or using covert communication methods like steganography to hide data within innocent-looking files. An attacker could also compromise hardware components or software updates during the supply chain process. Once these components are installed within the air-gapped network, hidden malware might activate and communicate with external parties. Note that APTs (Advanced Persistent Threats) in the past targeted highly secured and air-gapped networks, including USBStealer, Agent.BTZ \cite{gostev2014agent}, Stuxnet, Fanny, MiniFlame, Flame, Gauss, ProjectSauron, EZCheese, Emotional Simian, USB Thief, USBFerry, Brutal Kangaroo, Retro, PlugX, and Ramsay \cite{dorais2021jumping}. More recently, in August 2023, researchers at Kaspersky discovered another new malware and attributed it to the cyber-espionage group APT31, which targets air-gapped and isolated networks via infected USB drives \cite{Kaspersk37:online}.

\begin{figure}
	\centering
	\includegraphics[width=0.9\linewidth]{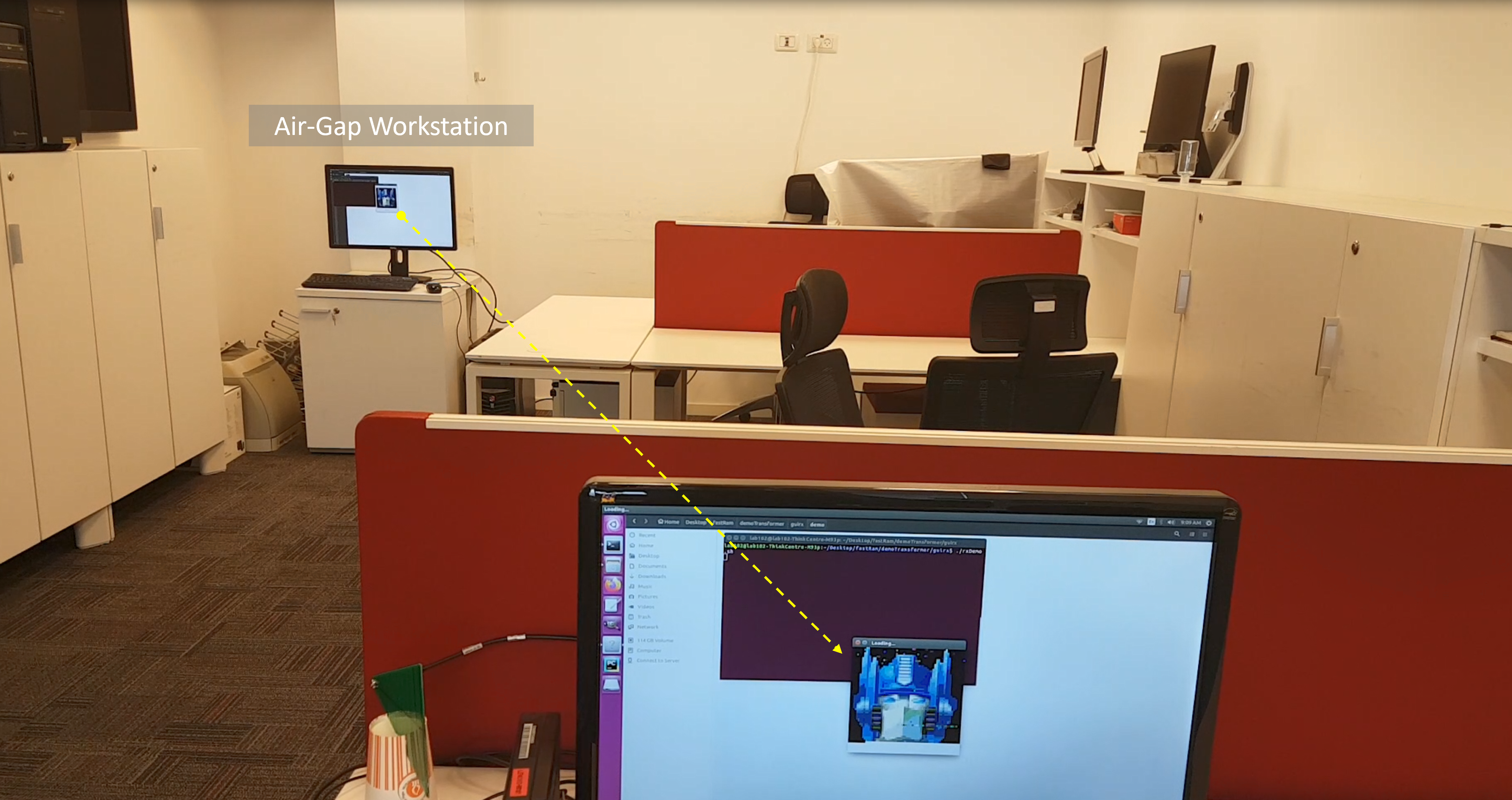}
	\caption{Attack demonstration. An air-gap workstation processes a secret image (Optimus Prime). The RAMBO covert channel attack transmits the image via electromagnetic waves. A remote attacker intercepts the information and recovers the secret image.}
	\label{fig:scenario}
\end{figure}

In the second phase of the attack, the attacker collects information, e.g., keylogging, files, passwords, biometric data, and so on, and exfiltrate it via the air-gap covert channel. In our case, the malware utilizes electromagnetic emissions from the RAM to modulate the information and transmit it outward. A remote attacker with a radio receiver and antenna can receive the information, demodulate it, and decode it into its original binary or textual representation. The attack scenario is illustrated in Figure \ref{fig:scenario}. The RAMBO malware within the infected air-gapped workstation (A) transmits sensitive images (Optimus Prime) using covert electromagnetic radiation from the RAM. A remote attacker intercepts the information and decodes the data.

\section{Related Work}
\label{sec:related}
Air-gap covert channels refer to a type of covert communication method that transfers information between two physically isolated systems or networks that are not directly connected through wired or wireless means. In the security research domain, air-gap covert channels are rooted in the idea that even systems disconnected from external networks might still communicate through unintended or concealed means. While the air gap is intended to prevent unauthorized data transfer, various techniques have been explored to bypass this isolation and create hidden communication channels. The main types of air-gap covert channels are acoustic, optical, thermal, and electromagnetic. In this paper, RAMBO covert channels are categorized as electromagnetic covert channels. In acoustic covert channels, systems might use ultrasonic sound waves inaudible to humans to transmit data between air-gapped devices. Specialized software or malware can encode data into sound signals picked up by a microphone on the receiving device \cite{Deshotels2014}\cite{de2022inkfiltration}. Previous work shows that attacks can exploit CPU and GPU fans \cite{guri2020fansmitter}\cite{guri2022gpu}, Hard-disk drives (HDD) \cite{guri2017acoustic}, CD/DVD noise \cite{guri2020cd}, and power-supply sound characteristics\cite{guri2021power} to modulate information over an air gap. Data can be encoded and transmitted using light signals, such as rapidly flashing LED lights or screen brightness changes \cite{guri2019brightness}. The receiving device might use a camera or light sensor to detect and decode signals. Previous work showed that attackers could exploit keyboards \cite{guri2019ctrl}, routers \cite{guri2018xled}, hard-disk drives (HDD) \cite{Guri2017}, and screen LEDs \cite{guri2019brightness} to modulate information over air gaps for long distances. In these cases, the receiver is a sensor or a camera. Attackers can transmit information by causing minor temperature fluctuations imperceptible to human senses but detectable by sensitive thermal sensors. E.g., the BitWhistper attack \cite{guri2015bitwhisper}, presented by Guri et al., shows that the CPU can generate thermal signals that nearby computers can sense to transfer data over air gaps. Electromagnetic emissions, often unintended byproducts of computational activities, can be modulated to encode data. These emissions can then be captured and interpreted by a receiver equipped with appropriate sensors. For example, malware might exploit electromagnetic emissions of a computer's central processing unit (CPU) to create a covert communication channel. Previous works focused on radio frequency covert channels including EMLoRa \cite{shen2021lora}, AirHopper \cite{Guri2014}, GSMem \cite{Guri2015}, Air-Fi \cite{guri2022air}, SATAn \cite{guri2022satan}, and Lantenna \cite{guri2021lantenna}.

\section{Transmission and Reception}
\label{sec:trans}
This section presents the implementation of the transmitter and receiver and the signal generation, data modulation, demodulation, and encoding and decoding schemes.

The RAM bus operates electrical lines or pathways that connect the CPU to memory modules. These pathways transfer data, instructions, and addresses between the CPU and RAM. The RAM bus includes various components \cite{romoddr}.

\begin{itemize}
	\item \textbf{{Data Bus.}} This is the portion of the RAM bus responsible for carrying the actual data being read from or written to memory. The data bus width determines the amount of data transferred simultaneously. For example, a 64-bit data bus can transfer 64 bits (8 bytes) of data in one operation.
	
	\item {\textbf{Address Bus.}} The address bus carries memory addresses that indicate the specific location in memory from which the CPU wants to read or write data. The address bus width determines the maximum amount of memory the CPU can access directly. For instance, a 32-bit address bus can address up to 4 gigabytes of memory.
	
	\item {\textbf{Control Lines.}} These lines carry control signals coordinating data transfer timing and sequencing. Control lines handle reading, writing, activating memory chips, and signaling when data is ready.	
\end{itemize}

When data is transferred through a RAM bus, it involves rapid voltage and current changes, mainly in the Data bus. These voltage transitions create electromagnetic fields, which can radiate electromagnetic energy through electromagnetic interference (EMI) or radio frequency interference (RFI). The frequency range of electromagnetic emanation from the RAM bus mainly depends on its specific clock speed, measured in megahertz (MHz) or gigahertz (GHz). This clock dictates how quickly data can be transferred between the CPU and memory. The emanation levels are influenced by other bus characteristics, including its data width, clock speed, and overall architecture. Faster RAM buses (e.g., DDR4 and DDR5) with wider data paths can lead to quicker data transfers with increased emissions.

\subsection{Signal Generation}
As explained above, when data is read from or written to memory, electrical currents flow through the RAM chips and the associated traces on the printed circuit board (PCB). These electrical currents generate electromagnetic fields as a byproduct, which radiates EM energy. To create an EM covert channel, the transmitter needs to modulate memory access patterns in a way that corresponds to binary data. For instance, they could alter the timing or frequency of memory access operations to encode information. The sender and receiver must establish rules that define how memory access patterns translate to binary values. For example, a reading or writing array to the physical memory for a specific timing interval might represent a '0' while another interval represents a '1'. The receiver detects and decodes the EM emissions caused by the modulated memory activity. This could involve sensitive radio frequency (RF) receivers or electromagnetic field sensors. 

\subsection{Modulation}
Algorithm 1 shows the signal generation with OOK (On-Off Keying) modulation, a basic form of digital modulation used in communication systems to transmit digital data over a carrier wave. In our case, the OOK modulation involves turning the carrier wave on and off to represent binary data, where the presence of the carrier wave generated by memory activity corresponds to one binary state ( "1"). The absence of the electromagnetic carrier wave (thread \texttt{sleep()}) corresponds to the other binary state ("0"). Note that to maintain the activity in the RAM buses, we used the MOVNTI instruction \cite{MOVNTI—S81:online}, which stands for Move Non-Temporal Integer. It performs a non-temporal store of integer data from a source operand to a destination memory location. This instruction is primarily associated with optimizing memory operations for certain types of data transfers, particularly in cases where the data is not to be reused immediately. Note that for the beginning of the transmission, we used the preamble sequence of \texttt{10101010}, allowing the receiver to be synched with the transmitter.

\begin{figure}
	\centering
	\includegraphics[width=0.7\linewidth]{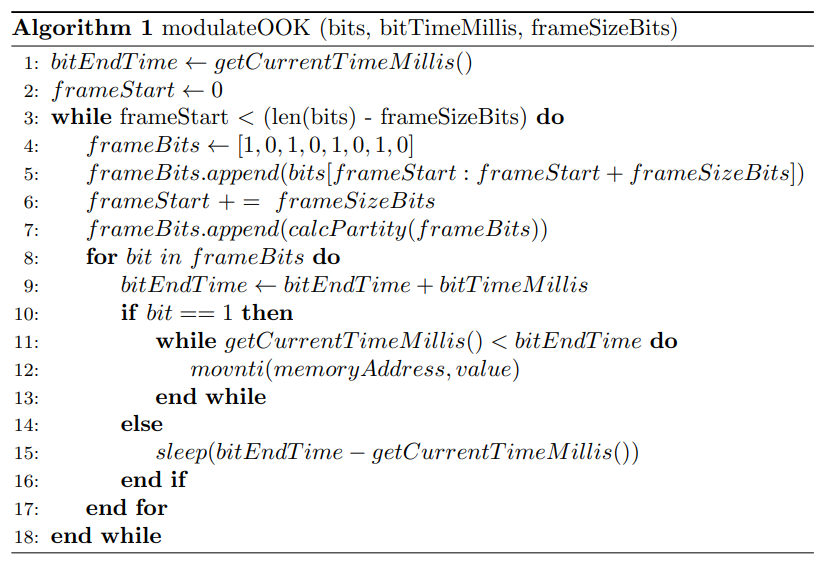}
	\caption{The RAMBO attack signal generation with OOK modulation }
	\label{fig:OOK}
\end{figure}

\subsection{Manchester Encoding}
For the fast transmission, we used the Manchester encoding. In this encoding, each bit of the binary data is represented by a transition or change in signal level within a fixed period.
Manchester encoding ensures a consistent number of signal transitions, making it useful for clock synchronization and error detection. The outline of our transmitter with Manchester encoding is presented in Algorithm 2.

\begin{figure}
	\centering
	\includegraphics[width=0.7\linewidth]{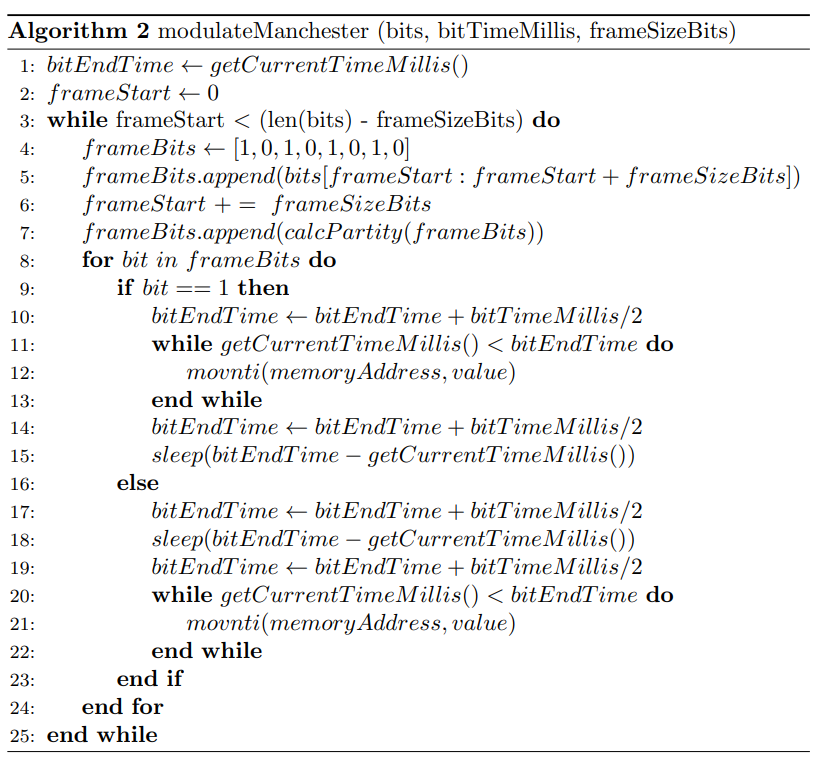}
	\caption{The transmission with Manchester encoding}
	\label{fig:MAN}
\end{figure}

\subsection{Demodulation and Framing}
\begin{figure}
	\centering
	\includegraphics[width=1\linewidth]{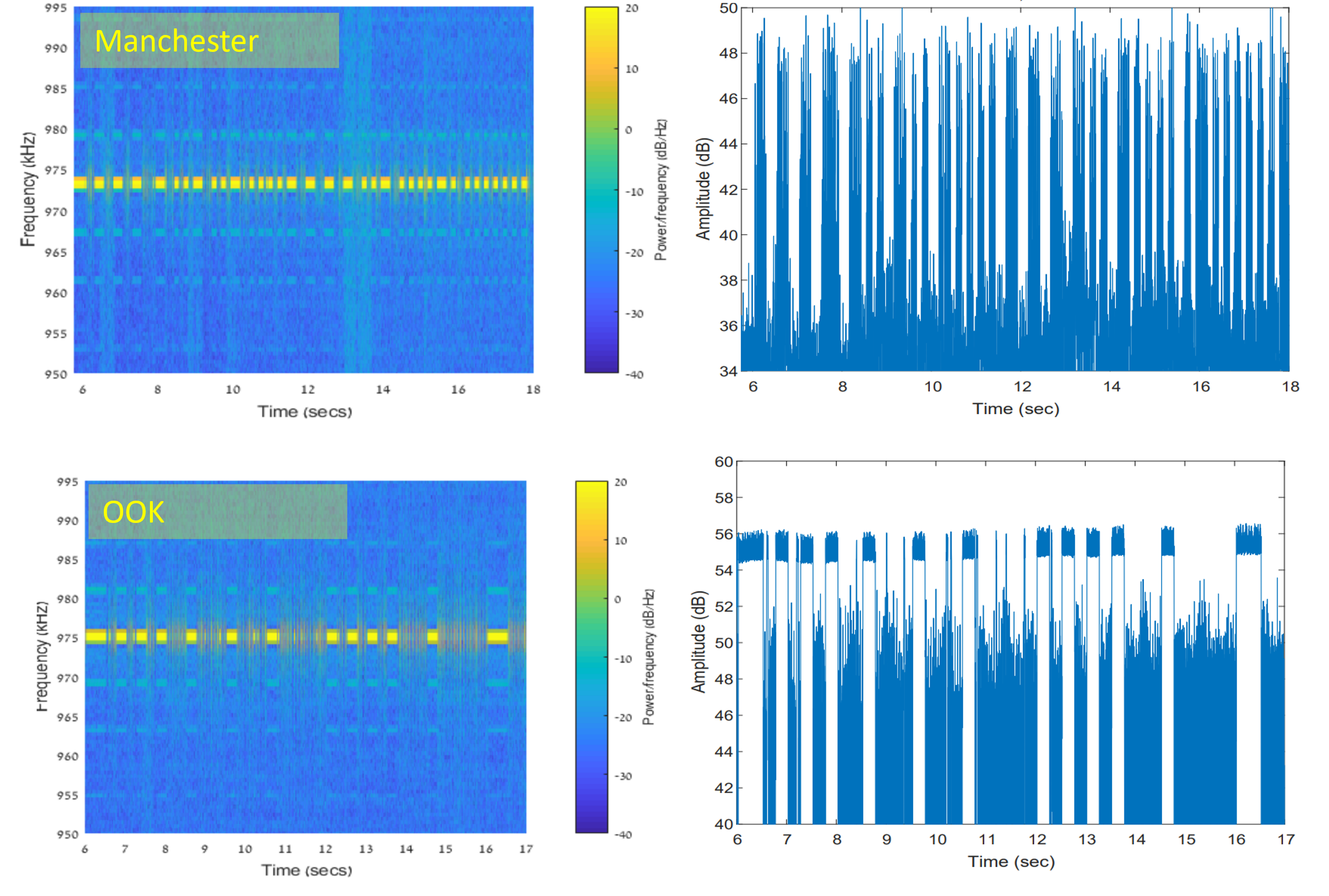}
	\caption{The signal of the word	`DATA' (\texttt{0x44 0x41 0x54 0x41}) in Manchester encoding (top) and OOK modulation (bottom).}
	\label{fig:DECOD}
\end{figure}

We encode the data with a frame consisting of an alternating sequence of eight alternating bits that represents the frame's beginning. Our demodulator is presented in Algorithm 3. 
Figure \ref{fig:DECOD} shows the spectrogram and waveform of the word `DATA' (\texttt{0x44 0x41 0x54 0x41}) transmitted in the Manchester encoding (top) and OOK modulation (bottom). Our analysis shows that the Manchester encoding is more relevant for the requirements of the RAMBO covert channel due to two main reasons; (1) the encoding aids in clock synchronization between the sender and receiver, and (2) the frequent transitions make it easier to detect errors caused by signal loss, interference, or distortion. However, it's important to note that Manchester encoding doubles the required bandwidth compared to direct binary encoding (e.g., the OOK), as each bit requires two signal transitions within the bit interval. This increased bandwidth requirement can be a drawback in some scenarios, especially for high-speed data transmission.

\begin{figure}
	\centering
	\includegraphics[width=0.7\linewidth]{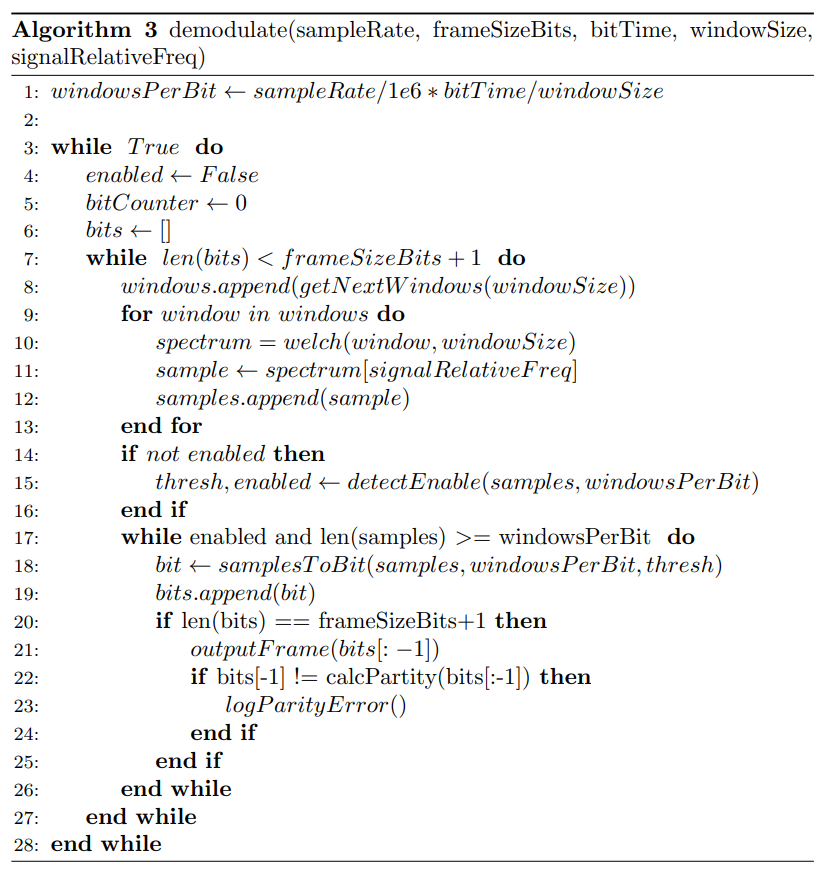}
	\caption{The demodulation algorithm}
	\label{fig:DEMOD}
\end{figure}

\section{Evaluation}
\label{sec:eval}
\begin{figure}
	\centering
	\includegraphics[width=0.5\linewidth]{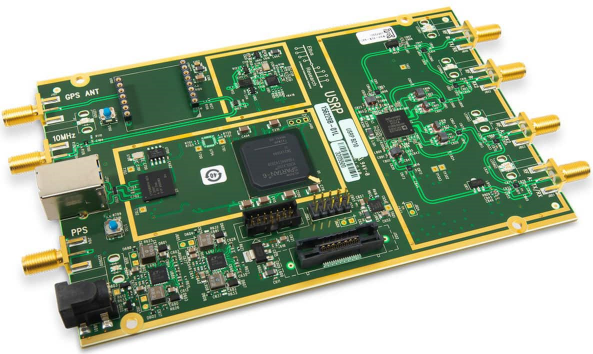}
	\caption{Ettus B210 Universal Software Radio Peripheral (USRP).}
	\label{fig:USRP}
\end{figure}

In this section, we present the evaluation of the covert channels. We tested three types of workstations. The PCs were all Intel i7 3.6GHz CPUs and 16GB of 2.133 - 2.400 GHz RAM. The PC ran Linux Ubuntu 18.04.6 LTS 64-bit. For the reception, we used the software-defined radio (SDR) Ettus B210, which is a specific model of the Universal Software Radio Peripheral (USRP) developed by Ettus Research, National Instruments (\ref{fig:USRP}). The B210 offers a wide range of capabilities for researchers, engineers, and enthusiasts working in wireless communication, radio frequency (RF) research, and signal processing. It covers a frequency range from 70 MHz to 6 GHz and supports sample rates of up to 61.44 MS/s (mega-samples per second). The USRP was connected to a small form factor NUC computer with 16 GB RAM running the C demodulator. It also ran MathWorks Matlab for signal processing and spectrogram visualization.

\subsection{Signal to Noise (SNR)}
\begin{figure}
	\centering
	\includegraphics[width=1\linewidth]{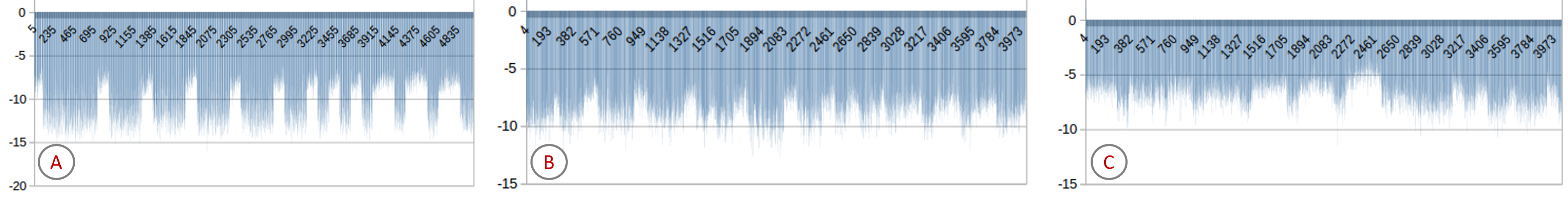}
	\caption{SNR with $t=250 ms$  (A), $t=100 ms$  (B), and $t=50 ms$ (C)}
	\label{fig:SNR}
\end{figure}

We evaluated the SNR levels at distances of 100 - 700 cm. Table \ref{tab:SNR} lists the average SNR levels. The SNR levels ranged from 38dB - 8 dB, which reflects the effective distance the covert channel can operate in this setup. Note that the SNR is also affected by the bit times. Figure \ref{fig:SNR} shows the thee different SNR with $t=250 ms$ (A), $t=100 ms $ (B), and $t=50 ms$ (C). As can be seen, the SNR is significantly affected by the bit time, with a differentiation of an average 7 dB between speeds with a shifting of 50 bit/sec.

\begin{table}[]
	\caption{The average SNR levels in a range of 50 - 700 cm}
	\centering
	\label{tab:SNR}
	\begin{tabular}{@{}|l|l|l|l|l|l|l|l|l|@{}}
		\toprule
		& d = 50 cm & d = 100 cm & d = 200 cm & d = 300 cm & d =  400 cm & d = 500 cm & d = 600 cm & d = 700 cm \\ \midrule
		Average & 38 dB     & 30 dB      & 27 dB      & 22 dB      & 17 dB       & 15 dB      & 12 dB  & 8 dB   \\ \bottomrule
	\end{tabular}
\end{table}

\subsection{Bitrates}
We evaluated the three speeds' effective bit rates and corresponding bit-error rates. Tables \ref{tab:t1}, \ref{tab:t5}, and \ref{tab:t10} shows the bit error rate (BER) values for $t=10 ms$, $t=5 ms$, $t=1 ms$, respectively. With a slow transmission rate ($t=10 ms$), A transmission is maintained at a distance of 700 cm away. With medium transmission rate ($t=5 ms$), A transmitted is maintained at a distance of 450 cm away and BER of 3\%-4\%. With a fast transmission rate ($t=1 ms$), a transmission is maintained at a distance of 300 cm away and BER of 2\%-4\%.

\begin{table}[]
	\caption{Transmission with $t=10$}
	\centering
	\label{tab:t10}
	\begin{tabular}{@{}|l|l|l|l|l|l|l|l|l|@{}}
		\toprule
		& d = 50 cm & d = 100 cm & d = 200 cm & d = 300 cm & d = 400 cm & d = 500 cm & d = 600  cm & d = 700 cm \\ \midrule
		PC-1 & 0\%       & 0\%        & 0\%        & 0\%        & 0\%        & 0\%        & 0\%         & 0\%        \\ \midrule
		PC-2 & 0\%       & 0\%        & 0\%        & 0\%        & 0\%        & 0\%        & 0\%         & 0\%        \\ \midrule
		PC-3 & 0\%       & 0\%        & 0\%        & 0\%        & 0\%        & 0\%        & 0\%         & 0\%        \\ \bottomrule
	\end{tabular}
\end{table}

\begin{table}[]
	\caption{Transmission with t=5}
	\centering
	\label{tab:t5}
	\begin{tabular}{@{}|l|l|l|l|l|l|l|@{}}
		\toprule
		& d = 50 cm & d = 100 cm & d = 200 cm & d = 300 cm & d =  400 cm & d = 450 cm \\ \midrule
		PC-1 & 0\%       & 0\%        & 0\%        & 1\%        & 2\%         & 4\%        \\ \midrule
		PC-2 & 0\%       & 0\%        & 0\%        & 0\%        & 2\%         & 3\%        \\ \midrule
		PC-3 & 0\%       & 0\%        & 0\%        & 0\%        & 3\%         & -          \\ \bottomrule
	\end{tabular}
\end{table}

\begin{table}[]
	\caption{Transmission with $t=1$}
	\centering
	\label{tab:t1}
	\begin{tabular}{@{}|l|l|l|l|l|@{}}
		\toprule
		& d = 50 cm & d = 100 cm & d = 200 cm & d = 300 cm \\ \midrule
		PC-1 & 0\%       & 0\%        & 0\%        & 4\%        \\ \midrule
		PC-2 & 0\%       & 0\%        & 0\%        & 3\%        \\ \midrule
		PC-3 & 0\%       & 0\%        & 0\%        & 2\%        \\ \bottomrule
	\end{tabular}
\end{table}

\subsection{Data Exfiltration}
Table \ref{tab:exfiltration} presents the time it takes to exfiltrate various types of information for three timing parameters ($t$). 
Keylogging can be exfiltrated in real-time with 16 bits per key (Unicode). A 4096-bit RSA encryption key can be exfiltrated at 41.96 sec at a low speed and 4.096 bits at a high speed. Biometric information, small files (.jpg), and small documents (.txt and .docx) require 400 seconds at the low speed to a few seconds at the fast speeds. This indicates that the RAMBO covert channel can be used to leak relatively brief information over a short period.   
\begin{table}[]
	\caption{Exfiltration time of various types of information with  RAMBO covert channel}
	\centering
	\label{tab:exfiltration}
	\begin{tabular}{@{}|l|l|l|l|l|@{}}
		\toprule
		Information                     & Size              & t = 10 ms & t = 5 ms  & t  = 1 ms \\ \midrule
		Keylogging                      & 16 bits (per key) & realtime  & realtime  & realtime  \\ \midrule
		4096 bit RSA key                & 4096 bits         & 41.96 sec & 20.48 sec & 4.096 sec \\ \midrule
		Biometric information           & 10000 bits        & 100 sec   & 50 sec    & 10 sec    \\ \midrule
		Password                        & 128 bits          & 1.28 sec  & 0.64 sec  & 0.128 sec \\ \midrule
		Small image (.jpg)              & 25000 bits        & 250 sec   & 125 sec   & 25 sec    \\ \midrule
		A textual document (.txt, .docx) & 40000 bits        & 400 sec   & 200 sec   & 40 sec    \\ \bottomrule
	\end{tabular}
\end{table}

\subsection{Faraday Shielding}
It is possible to block electromagnetic radiation from the computer using a specialized metal chassis built as a Faraday cage. 
The attenuation of a Faraday cage, which measures how effectively it blocks electromagnetic radiation, depends on various factors, including the frequency of the radiation, the conductivity of the cage material, and the thickness of the cage walls. 
The attenuation (A) of electromagnetic radiation by a conductive material like a Faraday cage can be approximated using Equation \ref{eq:att}. The attenuation factors are listed in Table \ref{tab:att}.

\begin{equation}
\label{eq:att}
	A = 10 \cdot \log_{10} \left( \frac{1}{1 + \left( \frac{\sigma d}{\mu f} \right)^2} \right)
\end{equation}

\begin{table}[]
	\caption{The computer chassis attenuation factors}
	\label{tab:att}
	\centering
	\begin{tabular}{@{}|l|l|@{}}
		\toprule
		Factor & Effect                                                      \\ \midrule
		$A$      & The attenuation in decibels (dB).                           \\ \midrule
		$\sigma$ & The conductivity of the material (siemens per meter, S/m).  \\ \midrule
		$d$      & The thickness of the material (meters).                     \\ \midrule
		$\mu$    & The permeability of the material (henries per meter, H/m).  \\ \midrule
		$f$      & the frequency of the electromagnetic radiation (hertz, Hz). \\ \bottomrule
	\end{tabular}
\end{table}

\begin{figure}
	\centering
	\includegraphics[width=1\linewidth]{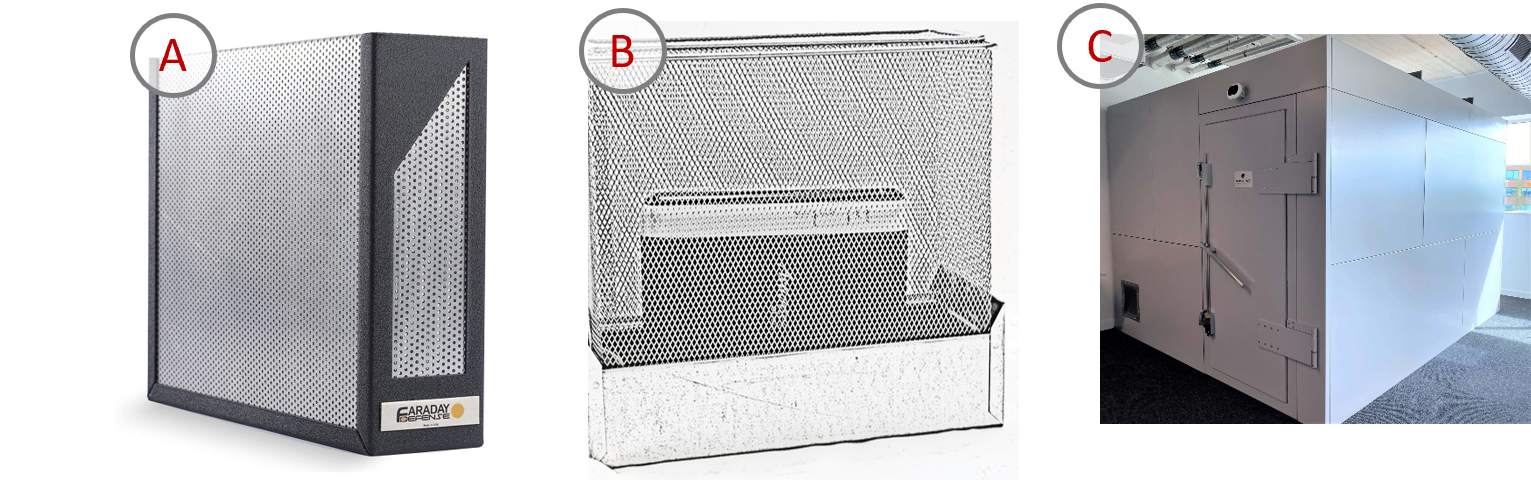}
	\caption{The PC sized Faraday enclosure (A), general size Faraday enclosure (B), and a Faraday room (C).}
	\label{fig:Faraday}
\end{figure}

We analyzed and measured the effect of the Faraday chassis on the RAMBO covert channel using copper foil. This has electric field high Shielding properties (above 100 dB) and magnetic fields. The copper standard width is 1 mm effectively blocks the EMR from the transmitting workstation. However, as noted in the following section, this solution is costly and can not be deployed widely. Another option is to use a Faraday room which is typically constructed using metal that can conduct electric currents. The primary purpose of a Faraday room is to create an electromagnetically isolated environment, which means that electromagnetic fields from external sources are significantly reduced or prevented from entering the enclosed space. Faraday enclosures are presented in Figure \ref{fig:Faraday} with a PC-sized Faraday enclosure (A), general size Faraday enclosure (B), and a Faraday room (C).

\subsection{Virtualization}
We evaluated the effectiveness of the covert channel when the transmitting code operates from within a virtual machine (VM). For the evaluation, we used VMWare workstation 16.2.4 running Linux Ubuntu 18.04.6 LTS 64-bit on host and guest machines. Our test shows that the low BER of below 1\%  was kept even when the code ran with a VM. However, it is essential to note that a massive workload in the host OS or memory activity in another guest OS might interrupt the signal generation conducted by the compromised virtual machine.

\subsection{Higher bit-rates}
\begin{figure}
	\centering
	\includegraphics[width=0.7\linewidth]{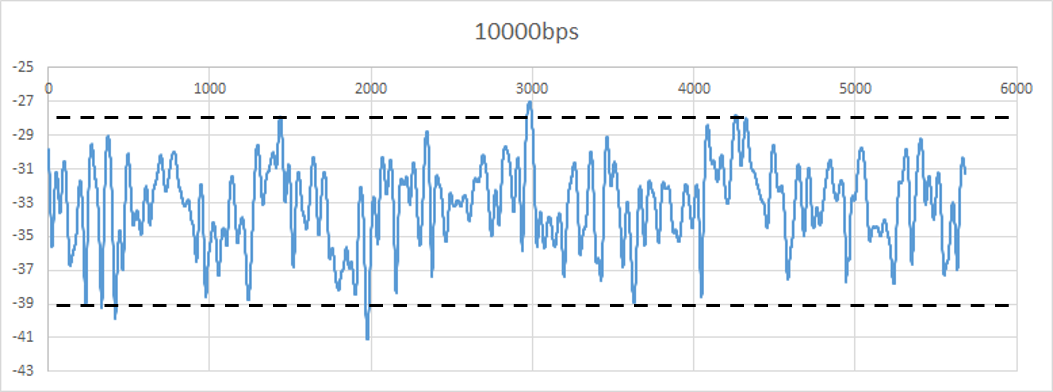}
	\caption{The transmission with 10000 bps.}
	\label{fig:10k}
\end{figure}

We tested the high bit rates of 5000 bps and above. Our evaluation shows that it is possible to demodulate the signal with mostly above 5\% BER, rendering this speed less effective. The main reason is the low SNR levels the fast signal generation yielded. Figure \ref{fig:10k} shows the waveform of the alternating short signal generated with 10000 bps. As depicted, the SNR is low (below 5\%) and causes high BER levels during the modulation.

\subsection{Frequency Ranges}
The electromagnetic emission from DDR RAM and other digital components can span a wide frequency range, including fundamental frequencies, subharmonics, and spurious missions. 
The central frequencies are the direct clock frequencies and their harmonics. For example, with DDR RAM operating at a clock frequency of 1.6 GHz (corresponding to DDR4-3200), we can observe emissions around 1.6 GHz, 3.2 GHz, and 4.8 GHz (3rd harmonic). It is important to note that DDR RAM modules emit frequencies that are not direct harmonics but are related to the clock frequency more indirectly; these can include subharmonics and other spurious emissions. We don't use these frequencies for the RAMBO covert channel. Our tests show that some systems use spread spectrum clocking to spread electromagnetic emissions across a broader range of frequencies. This technique can help reduce the concentration of emissions at specific frequencies, making it less likely to carry the modulated information. 

\section{Countermeasures}
\label{sec:counter}
Several defensive and protective countermeasures can be taken to defend against the proposed covert channel. 

\begin{itemize}
	\item \textbf{Zone restrictions.} The red-black separation concept involves creating a clear boundary or barrier between "red" and "black" components or environments to prevent unauthorized transfer of information from one domain to the other. This separation can be achieved through physical, logical, and procedural measures. In practice, defenders often use separate networks, hardware, and physical access controls to keep red and black systems physically separate from each other. There are several NATO and American standards, such as SDIP-27, AMSG, NSTISSAM, and ZONES, that mandate the segregation of areas that deal with the radiated electromagnetic, magnetic, optical, and acoustic energy of devices \cite{NSTISSAM75:online}. In this approach, radio receiver devices are eliminated from air-gapped computers or kept outside a specified radius of several meters away. The red-black separation concept may be applied in various domains, including military, intelligence, critical infrastructure, and organizations dealing with susceptible information \cite{NSTISSAM75:online}. In the context of the RAMBO attack, it can mitigate the risk of RAM leakage and unauthorized access by creating a clear separation between the two security domains.
	
	\item \textbf{Host intrusion detection systems (HIDS).} In this approach, we monitor the operating system's physical or virtual memory operations and detect suspicious operations. Such anomalies could be a process that abnormally reads and writes to memory regions. These are three different layers on which an intrusion detection system can operate. In this kernel-level approach, a driver/module is installed at the kernel level and continuously monitors the page access operations. Our experiment shows that all monitoring approaches imply high false positive rates. The main reason is that memory operations are always incurred by hundreds of threads in the OS, including the kernel level. Monitoring the analysis of these operations creates runtime overhead and leads to a high rate of false alarms.
	
	\item \textbf{Hypervisor-level memory access monitoring.} Because the hypervisor operates at a lower level of system control, it has visibility into the memory access patterns of the virtual machines it manages. This visibility allows the hypervisor to monitor memory access \cite{tang2018secure}. Technically, the hypervisor manages the virtual-to-physical memory mapping for each VM through memory page tables. An Extended Page Table (EPT) is a virtualization technology used in modern processors to enhance the performance of virtual machines (VMs) in a virtualized environment. EPT is specific to Intel processors and is equivalent to AMD's Nested Page Tables (NPT). By monitoring and controlling these page tables, the hypervisor can keep track of memory access patterns and perform memory isolation. Note that this approach was proposed by previous work to detect shellcode injection attacks and other types of vulnerability exploits. However, as the HIDS solution, it may lead to a high rate of false positives.

	\item \textbf{External radio monitoring.} Dedicated spectrum analyzers are specialized hardware devices designed to scan and analyze the radio frequency spectrum. They provide detailed information about signal strengths, frequency utilization, and interference sources. Spectrum monitoring refers to analyzing and observing the radio frequency (RF) spectrum used by wireless networks, e.g., those using the Wi-Fi standard. This monitoring helps understand the wireless communication environment's usage, interference, and overall health. It involves scanning and analyzing the different frequency channels within the RF spectrum to detect signals, identify sources of interference, and optimize the performance of wireless networks. In the context of a RAMBO attack, spectrum analyzers can provide visibility of the RF spectrum, showing signal strength across different frequency bands and detecting unintended transmission and covert channels.
	
	\item \textbf{Internal RAM jamming.} Another option is to interrupt the covert channel by applying random memory operations. Such jamming can be implemented as a user or kernel thread randomly interfering with memory and performing read-write operations. This approach has the main disadvantage of interfering with legitimate memory activities, resulting in significant overhead. In addition, internal jammer threads can be manipulated, bypassed, or terminated by malware with sufficient privileges running on a compromised computer.
	
	\item \textbf{External Electromagnetic jamming.} It is possible to defend against the covert channel using radio jammers, also known as signal jammers or RF (radio frequency) jammers. These devices are designed to interfere with or disrupt wireless communications by emitting radio frequency signals on the required frequencies used by the targeted communication systems. The goal of a radio jammer is to create a jamming signal that overwhelms and interferes with legitimate signals, rendering the communication systems ineffective or unreliable within the jamming area \cite{kasturi2020detection}. In the case of a RAMBO attack, the whole spectrum of DDR should be jammed. However, this approach requires dedicated external hardware transceivers and antennas, which are maintained in a secure area, which tend to pose another security threat.

	\item \textbf{Radio reduction/blocking Faraday enclosures.} A PC Faraday enclosure, also known as a Faraday cage or Faraday enclosure, is a shielded enclosure designed to block external electromagnetic fields and electromagnetic radiation from entering or leaving the enclosed space \cite{chapman2015mathematics}. This shielding helps protect sensitive electronic devices and equipment from electromagnetic interference (EMI) and prevents emitted electromagnetic radiation from leaking out and potentially interfering with other devices or systems. The Faraday enclosures will limit the leakage of radio frequencies of the RAMBO attack. However, the solution is costly and not applied on a broad scale.
\end{itemize}

Table \ref{tab:cnt} lists the countermeasures and their limitations.

\begin{table}[]
	\centering
	\caption{Defensive countermeasures}
	\label{tab:cnt}
	\resizebox{\columnwidth}{!}{%
		\begin{tabular}{@{}ll@{}}
			\toprule
			Solution                                               & Drawbacks                                         \\ \midrule
			Zone restrictions (red-black separation)                                 & Cost and space limitation                         \\
			Host intrusion detection systems (user/kernel) & High rates of false positive                      \\
			External electromagnetic spectrum monitoring                           & High rates of false positive                      \\
			Internal RAM operation jamming                               & Disruption of the RAM functionality and overhead          \\
			External radio jamming of RAM frequencies                  & Radio interference,  high cost, and power consumption \\
			Radio reduction/blocking Faraday enclosures                             & Cost and maintenance                               \\ \bottomrule
		\end{tabular}%
	}
\end{table}

\section{Conclusion}
\label{sec:conclusion}
We present an air gap covert channel attack that allows attackers to exfiltrate sensitive data from isolated computers. We show that malicious code in the infected computers can manipulate memory operations and generate radio signals from the memory buses. By precisely controlling the memory-related instructions, arbitrary information can be encoded and modulated on the electromagnetic wave. An attacker with a software-defined radio (SDR) can receive the information, demodulate it, and decide. We showed that this method could be used to exfiltrate arbitrary types of information, such as keystroke logging, files, images, biometric data, etc. We presented architecture and implementation, provided evaluation results, and discussed preventive countermeasures. With this method, attackers can leak data from highly isolated, air-gapped computers to a nearby receiver at a bit rate of hundreds bits per second.
%
%
%
\bibliographystyle{unsrt}

\end{document}